\documentclass[12pt,preprint]{aastex}

\usepackage{graphicx}

\newcommand{\be}{\begin{equation}}
\newcommand{\ee}{\end{equation}}
\newcommand{\bd}{\begin{displaymath}}
\newcommand{\ed}{\end{displaymath}}
\newcommand{\ba}{\begin{eqnarray}}
\newcommand{\ea}{\end{eqnarray}}

\shorttitle{3D-distribution of ISM}
\begin{document}

\date{}  
\title{3D Distribution of Molecular Gas in the Barred Milky Way}
\author{Martin Pohl}
\affil{Department of Physics and Astronomy,
Iowa State University\\
Ames, Iowa 50011-3160, USA}
\author{Peter Englmaier}
\affil{Institut f\"ur Theoretische Physik, Universit\"at Z\"urich,
  8057 Z\"urich, Switzerland}
\author{Nicolai Bissantz}
\affil{Fakult\"at f\"ur Mathematik, Ruhr-Universit\"at Bochum, 44780 Bochum, Germany}

\email{mkp@iastate.edu}

\begin{abstract}
We present a new model of the three-dimensional
distribution of molecular gas in the Milky Way Galaxy, based on CO line data.

{Our analysis is based on a gas-flow simulation {of the inner Galaxy 
using smoothed-particle hydrodynamics (SPH) using} a realistic barred
gravitional potential derived from the observed
COBE/DIRBE near-IR light distribution. The gas model prescribes the
gas orbits much better than a simple circular rotation model and is
highly constrained by observations, but it cannot {predict} local details.
In this study, we provide a 3D map of the observed molecular gas distribution using the
velocity field from the SPH model.
}
A comparison with studies of
the Galactic Center region suggests that the main structures are reproduced but somewhat stretched
along the line-of-sight, probably on account of limited resolution of the underlying SPH simulation.
The gas model will be publicly available and may prove useful in a number of applications, 
among them the analysis of diffuse gamma-ray emission as measured with GLAST.
\end{abstract}

\section{Why another deconvolution of gas data?}
Models of the distribution of interstellar gas reflect the structure of the 
Milky Way Galaxy and have therefore considerable merit in themselves. 
In addition, they are also highly valuable for a
variety of other applications, among them the physical analysis of diffuse Galactic
gamma-ray emission \citep{ber93,hu97,pe98}. 
Knowledge of the gas distribution is essential for studies of the 
large-scale cosmic-ray
distribution in the Galaxy, as well as for investigations of small-scale variations
in the density and spectrum of cosmic rays.
The upcoming launch of GLAST, a GeV-band
gamma-ray observatory of unprecedented sensitivity, makes it desirable to have 
an up-to-date
model of the three-dimensional distribution of the ISM. There are two main 
reasons why a new study would be required: 
The various components of interstellar gas
are traced by their line emission, and the quality of the line data available today
is much higher than it was in the Nineties. Also, it is now well established that 
the Galaxy contains a central bar \citep[e.g.][]{Babu+Gil05,ben05}, which 
causes non-circular motion of interstellar gas in the inner Galaxy, thus changing the kinematic
relation between the location on the line-of-sight and the velocity relative to the 
local standard of rest (LSR).

Here we report results for the deconvolution of CO$_{1\rightarrow 0}$ data 
for the entire Galactic plane \citep{dame01}. For that purpose we use a 
gas-flow model derived from 
smoothed particle hydrodynamics (SPH) simulations in gravitational
potentials based on the NIR luminosity distribution of the bulge and disk
\citep{biss03}. Besides providing a more accurate picture of cloud orbits in the inner
Galaxy, {a} fundamental advantage of this model is
that it provides kinematic resolution toward the Galactic Center, in
contrast to standard
deconvolution techniques based on purely circular rotation \citep{ns06}. 
\citet{saw04} used OH-absorption data in comparison with 
CO emission lines to infer the distribution of molecular gas in the inner few hundred 
parsecs, and found it strongly influenced by the bar. Our model should incorporate 
the imprint of the bar over the entire inner Galaxy. {{Therefore,}
our result does not suffer from a strong finger-of-god effect
like in the classical paper by \cite{oort58}. }

{Because any deconvolution will introduce artefacts, we test our 
procedure on simulated line spectra, which allows us to identify artefacts that are present 
in the final gas model. We investigate three different gas-flow models for the inner Galaxy, one 
of which is intentionally distorted so it no longer corresponds to a SPH simulation that has been 
{adapted} to gas data.}

In the Galactic-Center region,
for which studies with alternative methods like
OH absorption have been performed, we find the gas distribution in
our model generally consistent with those earlier results, provided one accounts for the 
existence of barred gravitational potential. On account of limited resolution both in 
the gas flow model and the deconvolution the central molecular zone appears somewhat stretched
along the line-of-sight, though.

\section{The method}
\subsection{The CO data}
The CO$_{1\rightarrow 0}$ emission line is {the best available} tracer of molecular gas, even 
though the exact relation between the integrated line intensity and the column density
of molecular gas, usually referred to as the X-factor, is known to vary with
Galactocentric radius and metallicity \citep{sod95,ari96,oka98,strong04,nakagawa}.
The X-factor can be determined at specific locations through the line signal of
CO with rare isotopes of either carbon 
or oxygen using various assumptions for the radiation transport
\citep{dick78,dahm98,wall06} or through absorption measurements of H$_2$ and CO in
the UV \citep{burgh07}. 

We use the composite survey of \citet{dame01}, which comprises more than 30 individual 
surveys of CO$_{1\rightarrow 0}$ emission that together cover the entire Galactic Plane.
The data were taken with the CfA 1.2-m telescope and a similar instrument in Chile.
The angular resolution is about 1/6 of a degree, and the sampling 
is slightly better than that with 1/8 of a degree. The velocity sampling is 1.3~km/s
and the rms noise is around 0.3~K per channel, but varies slightly over the Galactic
Plane. The advantage of this survey lies in its sensitivity, the sampling, and 
the uniformity.

CO surveys have been conducted with significantly higher angular resolution, 
for example the 
FCRAO Outer Galaxy Survey \citep{heyer98}, the
Massachusetts-Stony Brook Galactic Plane CO Survey \citep{Clemens86}, or the NANTEN 
Galactic Plane Survey. These surveys either cover only a small part of the sky or are 
significantly undersampled, thus somewhat compromising their applicability in
studies of the large-scale distribution of molecular gas in the Galaxy. Also, 
considering the {width of the} point-spread function
and the photon statistics at high 
energies, the effective angular resolution of GLAST is not better
than that of the CfA survey, and thus CO data of higher angular resolution may not
be needed. We have therefore decided to solely use the CfA survey
with a sampling of 1/8 of a degree. By applying the appropriate smoothing
we have verified, that the publicly
available high-resolution surveys are perfectly consistent with the lower-resolution
CfA survey, so by using the CfA survey we have not lost any
significant information other than the detailed distribution on scales below 1/6 of a degree.

\subsection{The Galactic bar}
\label{section-bar}
While bars were clearly observed in other galaxies, absorption of visible light
by dust has for a long time impeded searches for similar structures in our Galaxy. 
The availability
of sensitive infrared detectors in recent decades has finally permitted increasingly
accurate studies of the structure of inner Galaxy. Today, the observational evidence
for the existence of a Galactic bar is very strong \citep{ger02}, but some uncertainty
remains concerning the characteristics of that bar. As an example, using Spitzer data
\citet{ben05} find a bar with half-length of $R_{\rm bar}\simeq 4$~kpc
(for a GC distance of 8~kpc) at an angle of $\phi\simeq 45^\circ$, 
whereas the NIR photometry data of \citet{Babu+Gil05} suggest 
$R_{\rm bar}\simeq 2.5$~kpc and $\phi\simeq 22^\circ$. Using COBE/DIRBE L-band data
and giant-star counts \citet{bg02} determine the bar {to be oriented at an angle
$\phi\simeq 23^\circ$} with spiral arms emerging at $R=3.5$~kpc.

{Based on their earlier analysis of the COBE L-band data \citep{biss97},
\citet{engl99} have calculated the resulting gravitational potential and 
modelled the gas flow for the Milky Way inside the solar circle using
smoothed particles hydrodynamics in those potentials. It is worth noting,
that this model is non-parametric and has virtually no free parameters, {except} the bar
orientation is not tightly constrained by the observations {of
NIR-light, the microlensing event rate, the red clump giant distribution, and
the} CO kinematics. Later, in a
refined analysis, also the spiral arm pattern {was taken} into
account \citep{biss03}.}
{To select the best fitting models
of the gas flow, \citet{biss03} compared simulated longitude-velocity diagrams with the main features 
of observed CO emission for a certain range in Galactic longitude.}

Here we use {the velocity field from} their gas flow models
instead of a {simple} circular rotation curve to determine a distance-velocity
relation that will allow us to find the location of molecular gas as traced by
CO$_{1\rightarrow 0}$ line emission. It is important to note that the non-circular flows
imposed by the bar provide kinematic resolution even toward the Galactic Center
on account of the radial motion of gas. {We use three different 
{velocity-field models} in this work. The first is the standard model of
\citet{biss03}, which is based on a bar inclination angle $\phi=20^\circ$. As alternative we test
a model with $\phi=30^\circ$ that, according to \citet{biss03}, can also reproduce the main features in 
position-velocity diagrams of CO line data. The third model is the standard model rotated by $20^\circ$, so
the bar would make an angle of $\phi=40^\circ$ to the line-of-sight and line-of-sight velocities
are no longer in accord with the observed velocities of gas. The purpose of that third model is to show
the effect of an ill-fitting gas-flow model, and the deduced gas distribution in the 
Galaxy should be seriously distorted in this case.}

\begin{figure}
  \plotone{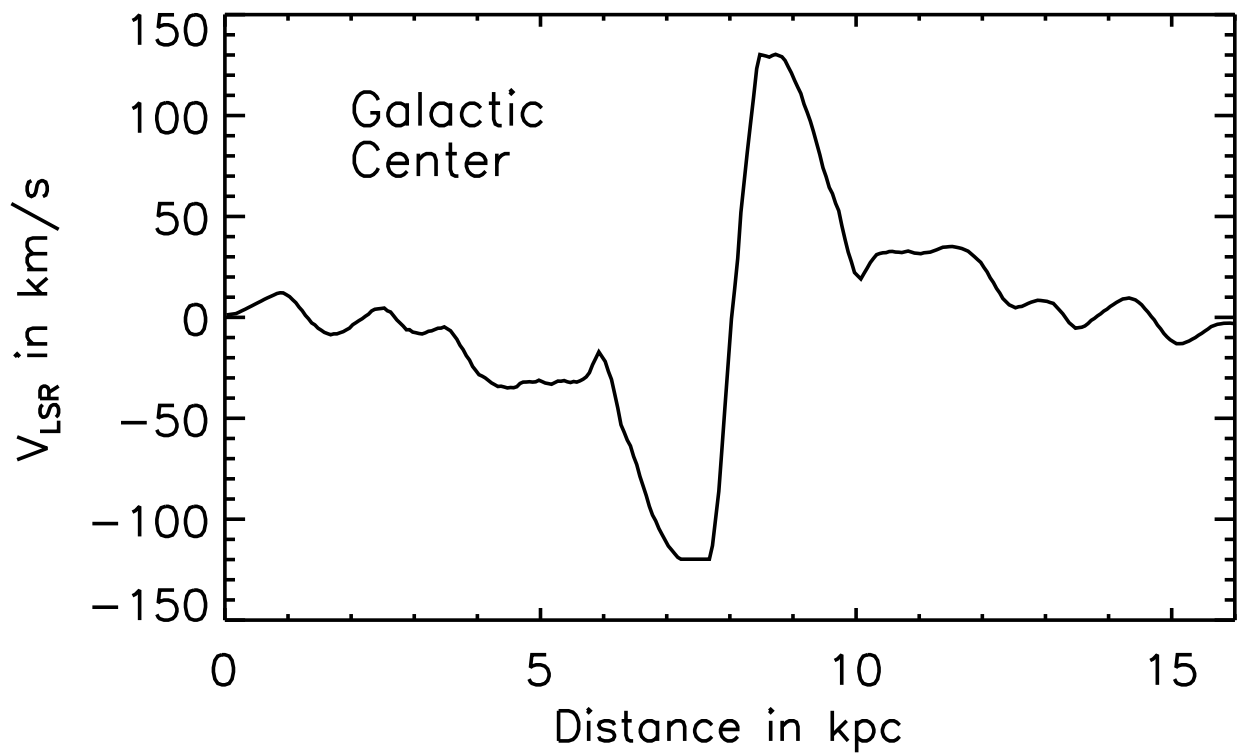}
\caption{The line-of-sight velocity of interstellar gas as a function of distance for 
the Galactic Center direction, based on the standard gas flow model of \citet{biss03}. The model
successfully predicts large radial velocities, but offers a large variety of possible 
distance solutions near zero velocity.}
\label{f1}
\end{figure}

\begin{figure}
  \plotone{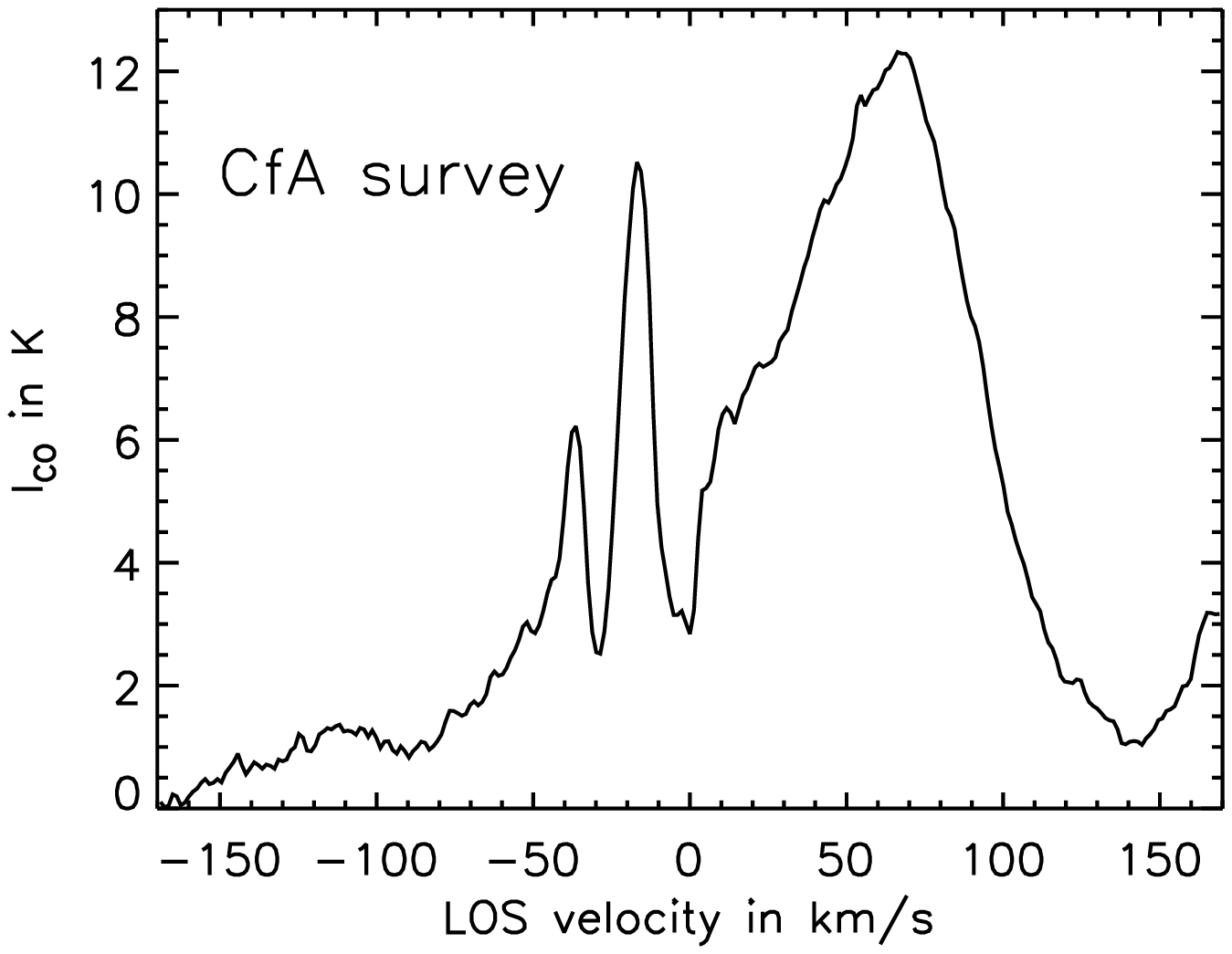}
\caption{The CO$_{1\rightarrow 0}$ line spectrum for the Galactic Center direction as
given by the CfA data cube. For most of the line signal distances can be found, but some
intensity at forbidden velocities remains.}
\label{f2}
\end{figure}

Fig.~\ref{f1} shows the line-of-sight velocity for the standard flow model
as a function of distance toward the Galactic Center.
For comparison, Fig.~\ref{f2} gives the CO line spectrum for the same line-of-sight.
For the bulk of the line signal at velocities around 50~km/s we now find two possible 
distance solutions, one near 8~kpc and the other one close to 10~kpc. 
To be noted from Fig.~\ref{f1} is also that for many velocities
we find a multitude of possible distances, e.g. eight different solutions near zero
velocity for the line-of-sight toward the Galactic Center, 
not just two as in the case of purely circular rotation. 
Also, the gas flow model does not {fully cover the observed range of velocities,}
as the spectrum
in Fig.~\ref{f2} shows a line signal of about 1~K at -150~km/s and of 2~K at 160~km/s,
which is far beyond the range of velocities for which distance solutions exist.


\subsection{Deconvolution technique}
Many quantities are practically expressed in galactocentric cylinder coordinates 
$(r,\phi,z)$, but transformations are easier to follow in galactocentric Cartesian 
coordinates $\vec r$.
In those coordinates any location is described by a vector
$\vec r=(r\,\cos\phi,r\,\sin\phi,z)$, and the sun is located at $\vec r_s =(R_0,0,z_0)$,
where \citep[e.g.][]{joshi}
\be
R_0=8\ {\rm kpc}\qquad\qquad z_0=15\ {\rm pc} \label{e1}
\ee
Let us write the distance vector $\vec y = \vec r-\vec r_s$ in the form
\be
\vec y =(-D\,\cos l\,\cos b, -D\,\sin l\,\cos b,D\,\sin b)=
(-P\,\cos l,- P\,\sin l,Q)\ \label{e2}
\ee
where we used heliocentric Galactic coordinates, $D$ is the true distance, and $P$ is the distance
as projected on the Galactic plane. For a given CO line spectrum 
we know the direction in Galactic coordinates $l$,$b$ and can relate
$r$ to the line-of-sight velocity. The solution for $P$ then is
\ba
P&=&R_0\,\cos l + \sqrt{r^2-R_0^2\,\sin^2l}\qquad\qquad r\ge R_0\\
P&=&R_0\,\cos l \pm \sqrt{r^2-R_0^2\,\sin^2l}\qquad\qquad r\le R_0
\label{e3}
\ea
where always 
\be
r\ge R_0\,\vert \sin l\vert
\ee
and also
\be
D\,\cos b=P\qquad\qquad z=z_0 + P\,\tan b
\label{e4}
\ee
Throughout this paper we assume that the gas flow in the Galaxy is independent of $z$,
so the flow pattern known for the mid-plane also applies at any height above the plane. 
This approximation should not cause much error because molecular gas is usually found close
to the mid-plane. {This assumption holds when $|z|<< r$, i.e. it
  breaks down very close to the {Galactic Center.}}

If we had a purely circular gas flow with rotation curve $V(r)$, then
the uncorrected radial velocity would be
\be 
V_{\rm LSR}(l,b,P) = \left({{R_0}\over r}\,V(r) -V(R_0)\right)\,\sin l\,\cos b 
\label{e5}
\ee
where Eq.~\ref{e3} is used to link the projected distance $P$ to the Galactocentric radius
$r$. The proper motion of the sun relative to the local standard of rest
has a line-of-sight component of \citep{db98}
\be
V_{\rm LSR,\odot} = \left(10.\,\cos l\,\cos b + 5.2\,\sin l\,\cos b+7.2\,\sin b\right)
\ {\rm km\,s^{-1}}
\label{e6}
\ee
which we can account for by subtracting it from the radial velocity.
The corrected, effective velocity then is
\be
V_{\rm eff} (l,b, P) = V_{\rm LSR} (l,b)-V_{\rm LSR, \odot}
\label{e7}
\ee
We use Eqs.~\ref{e5} and \ref{e7} to estimate the gas velocity for all Galactocentric
radii larger than $R_0$, i.e. outside the solar circle, where the gas orbits
are assumed circular with constant velocity $V=V_0=210$~km/s. The actual rotation
velocity of gas in the outer Galaxy depends somewhat on the
choice of $R_0$, the distance to the Galactic Center. A flat rotation curve with $V_0=210$~km/s
is a reasonable compromise between faster rotation for larger $R_0$ \citep{levine,mcclure}
and slower rotation for small $R_0$ \citep{olling}. Inside the solar circle 
we need to allow for non-circular rotation  and therefore
the gas velocities are given by the flow model described in subsection \ref{section-bar}.
A linear transition is used to match the gas flow model for the inner and outer Galaxy between
7~kpc and 9~kpc in Galactocentric radius. The flow model of \citet{biss03} also has
a few small data holes that are patched by linear interpolation.

While the bar model provides kinematic resolution toward the Galactic Center, we still face
a lack of resolution in the direction of the anticenter, because purely circular rotation
is assumed to apply for $r > R_0$.
Therefore the rotation curve is used only for $\vert b\vert \le 5^\circ$ and 
$\vert l\vert \le 165^\circ$. Towards the anticenter, where the kinematic resolution vanishes, 
we interpolate the distribution of gas between those derived in 10-degree windows centered on
$l=160^\circ$ and $l=200^\circ$ and use that as a probability function according
to which the actually measured line signal is distributed. 
At high latitudes ($\vert b\vert \ge 5^\circ$) the signal is distributed according to the 
distance distribution derived for $\vert b\vert \le 5^\circ$ as a prior, weighted by a
Gaussian of the height above the mid-plane, $z$, for the distance and latitude in question.

\begin{figure}
\epsscale{0.95}
  \plotone{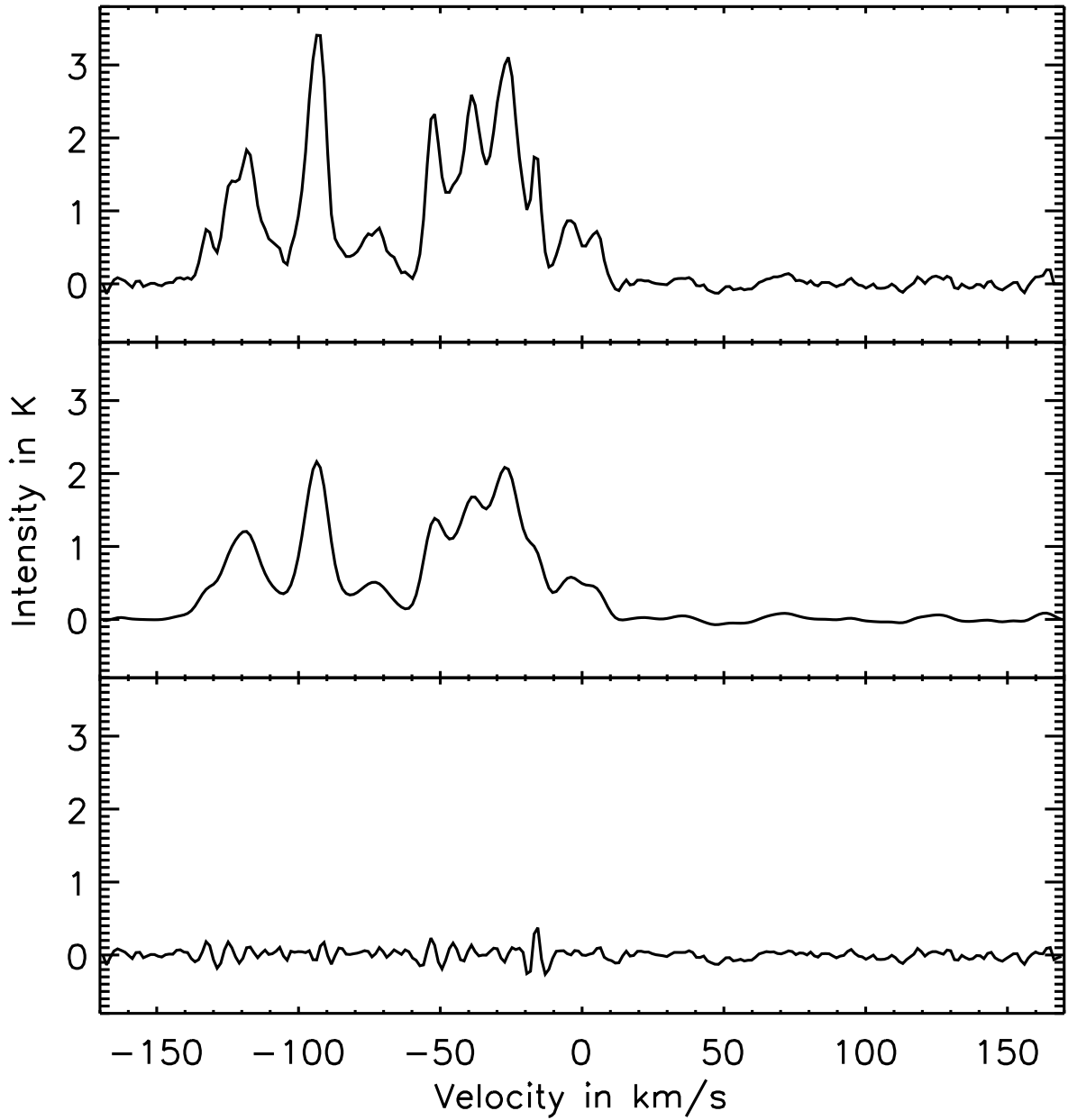}
\caption{The CO$_{1\rightarrow 0}$ line spectrum in the direction $l=340^\circ$ and $b=0^\circ$
at different stages of the deconvolution. The top panel gives the original spectrum as reported
by \citet{dame01}. The middle panel shows that spectrum after convolution with a Gaussian, which we use 
find the intensity and velocity of the peak emission, assuming true emission lines are significantly
broader than the velocity bins of the CO spectrum. The bottom panel displays the remaining spectrum
after termination of the deconvolution, which ideally should contain
only noise. {The residual spectrum shown in the bottom panel is typical. If the velocity dispersion 
(Eq.~\ref{e8}) were chosen too large, negative spikes could appear that arise from over-subtraction
of line wings.} 
}
\label{f8}
\end{figure}

The final data cube of deconvolved molecular gas will give the gas density in 
bins of 100~pc length for the line-of-sight distance, {assuming a nominal X-factor
$X={2.3\cdot 10^{20}\ \rm mol./cm^2/K/(km\,s^{-1})}$.}
The actual deconvolution uses distance
bins of 50~pc length, though, and follows an iterative procedure. 
The internal velocity dispersion of individual gas clouds is determined from the 
profiles of narrow lines as
\be
\sigma_{\rm CO}=3\ {\rm km/s}
\label{e8}
\ee
This single-cloud velocity dispersion is small, but still
well in the range of those derived in other studies \citep[e.g.][]{malh94}.
In the Galactic-Center region the velocity dispersion is expected to be higher than that
\citep{dahm98}. 
Within the central kiloparsec we therefore use
\be
\sigma_{\rm CO} (r\le 1\ {\rm kpc})=5\ {\rm km/s}
\label{e9}
\ee
The actual deconvolution consists of many steps, each of which is supposed to involve an
individual gas cloud or part thereof. For that purpose the
CO line spectrum is convolved with a Gaussian with half the velocity 
dispersion of individual gas clouds. The value and velocity of the peak in
the convolved spectrum is determined, which is less influenced by noise 
than if determined through the raw spectrum. {Tests have shown that
the deconvolution tends to break the total line signal into a few blobs on the 
line-of-sight, if we place the full peak line signal at the distance corresponding
to its velocity, so an iterative process is required. Providing both good computational
speed and accuracy,} a Gaussian with 20\% of
that peak value
(or the remaining velocity-integrated intensity, whichever is smaller)
and a dispersion as given by Eqs.~\ref{e8} or \ref{e9}, i.e.
\be
I(v)={{\delta W_{\rm CO}}\over {\sqrt{2\pi}\,\sigma_{\rm CO}}}\,
\exp\left[-{{(v-v_0)^2}\over {2\,\sigma_{\rm CO}^2}}\right]\ ,
\label{e9b}
\ee
is subtracted from the original spectrum and the corresponding
$\delta W_{\rm CO}$ value is added to the vector that represents the density distribution 
along the line-of-sight, thus ensuring that negative fluctuations are not propagated to 
the density distribution. This procedure is repeated until the velocity-integrated 
intensity in the remaining CO
line spectrum is less than a specified value, here $1\ {\rm K\,km\,s^{-1}}$. The
line spectrum that remains when the deconvolution has terminated should contain only noise.
Figure~\ref{f8} shows for a specific line-of-sight the original CO line spectrum in comparison
with the convolved spectrum, that is used to find the true peak velocity, and the remaining line 
spectrum at the end of the deconvolution. To be noted from the figure is that the 
remaining spectrum does indeed appear to be essentially noise. In the present case the
remaining integrated line signal is $0.94\ {\rm K\,km\,s^{-1}}$. It is possible that some of 
that remaining signal is true emission, and more sensitive observations of the outer Galaxy 
have in fact found CO line emission at lower level \citep{nakagawa}, 
but here we have no information as to the velocity (and hence
the distance) that this emission should be attributed to.
This translates to a systematic uncertainty 
in the reconstructed column density of molecular gas that can be estimated as
\be
\Delta N_{H_2} = X\,\Delta W_{\rm CO}\simeq 4\cdot 10^{20}\ {\rm atoms\,cm^{-2}}
\label{e9c}
\ee
for a standard X-factor. Only at high latitude, where the
gas column density is low, will this uncertainty have significance compared with uncertainties in 
the X-factor, limited coverage, or measurement uncertainties. In terms of Galactic diffuse 
$\gamma$-ray emission 
above 100~MeV the associated uncertainty is 30\% of the extragalactic diffuse background
as measured with EGRET \citep{sree}.

In each iteration step we determine
the 8 kinematically best-fitting locations using distance bins of 50 pc length
to account for the multitude of possible distance solutions. 
The line signal is distributed among these eight solutions with weights that are
determined by three factors: first a Gaussian function in the separation from the midplane,
\be
w_z= \exp\left(-{{(z-z_c)^2}\over {2\,\sigma_z^2}}\right)\ ,
\label{e10}
\ee
then a Gaussian function with a HWHM of 8.3~kpc in Galactocentric radius to limit 
the placing of gas at large
radii on the far side of the Galaxy, which is often kinematically allowed, but unlikely.
The last factors reflects the Jacobian $\vert dv/dD \vert$ that arises from transforming
a differential in velocity into a differential in line-of-sight distance, $D$. In that 
Jacobian we have to account for the binning in both velocity and distance. The 
data cube of the CO line spectrum provides us with the average intensity per velocity interval
$\delta v = 1.3$~km/s. The Jacobian translates this into an associated distance interval
$\Delta_v D$, over which the line signal should be distributed,
\be
\Delta_v D= \bigg\vert {{dv}\over {dD}} \bigg\vert^{-1}\,\delta v\ .
\label{e10a}
\ee
The distance itself is binned with $\delta D=50$~pc. If $\Delta_v D \ge \delta D$,
which is the standard case, then
the signal must be distributed over neighboring bins, each of which receives a fraction
$\delta D/\Delta_v D$ of the total signal. Our accounting for 8 possible distance solutions
at each iteration step together with the usually large number of iteration steps
ensures that in this case part of the line signal is indeed attributed to
the neighboring distance bins. If $\Delta_v D \le \delta D$,
then the signal must nevertheless be distributed over the entire distance bin. In total 
we can define weight factors representing the Jacobian as
\be
w_J={{\delta v}\over {\delta D}}\,\cdot\,\cases{1\qquad&{\rm for}\ $\Delta_v D \le \delta D$\cr
 & \cr
 {{\delta D}\over {\Delta_v D}}&{\rm for}\ $\Delta_v D \ge \delta D$}
\label{e10b}
\ee
Gas with forbidden velocity is placed in the distance bins with the best matching velocity, 
except toward the inner Galaxy ($\vert l\vert \le 20^\circ$) where for a velocity offset 
of more than 10~km/s to the nearest allowed velocity
we accept only distance bins in the Galactic Center region. 
Finally the line-of-sight distribution of gas is reduced to a resolution of 100~pc.

In Eq.~\ref{e10} we must account for an increase in the thickness of the gas disk with
Galactocentric radius. While the variations of $\sigma_z$ appear small within
the solar circle \citep{malh94,ns06}, a substantial flaring of the molecular gas disk is
observed in the outer Galaxy \citep{wout,bm98}. An analytical approximation to the various 
results reported in the literature is given by
\be
\sigma_z=60-50\,{r\over {R_0}}+60\,\left({r\over {R_0}}\right)^2 \ {\rm pc}
\label{e11}
\ee
Warping of the molecular disk appears insignificant within the solar circle 
\citep{malh94,ns06}, and therefore we neglect it altogether. In the outer regions of the Galaxy
the warp in the molecular disk is assumed identical to that of the H$I$ disk and
is given in good approximation by \citep{bm98}
\be
z_c=(1,000\,{\rm pc})\,x\,\sin\phi + (300\,{\rm pc})\,x^2\,(1-\cos(2\,\phi))
\label{e12}
\ee 
where 
\bd
x={{r-(11\,{\rm kpc})}\over {6\,{\rm kpc}}}\quad{\rm and}\ r\ge 11\,{\rm kpc}
\ed
{Finally, to somewhat alleviate the near-far ambiguity toward the inner Galaxy, we first deconvolve 
the spectra for Galactic latitudes $\vert b\vert \ge 1^\circ$. We then average the
deduced gas density in the near region over $0.9375^\circ < \vert b\vert < 1.5625^\circ$
and interpolate the result for latitudes $\vert b\vert < 1^\circ$, where they are used 
as an estimate for the minimum gas density in the near region. This procedure avoids the placing 
of the signal from nearby gas clouds to large distances, where they would correspond to a substantial 
surface mass density.}

\section{Test on simulated data}
{To test the deconvolution procedure and estimate the nature and strength of deconvolution 
artefacts, we have simulated a CO line data set using} {an artificial}{model gas 
distribution that consists of a bar and two spiral arms as shown in the top panel of
figure \ref{f9}. One should note that the simulated line data set is 
based on the same gas flow model and intrinsic line widths that are used during the deconvolution, so 
this test will not show artefacts that arise from imperfections of the flow model. Its main purpose is 
to identify the noise level and those characteristics in the final deconvolved gas distribution that are 
likely not real, {{as for example caused by distance ambiguities or}
{\em velocity-crowding}, i.e. small $|V/dD|$, where the inversion of the distance-velocity
relation is uncertain}.
\begin{figure}
\epsscale{0.5}
   \plotone{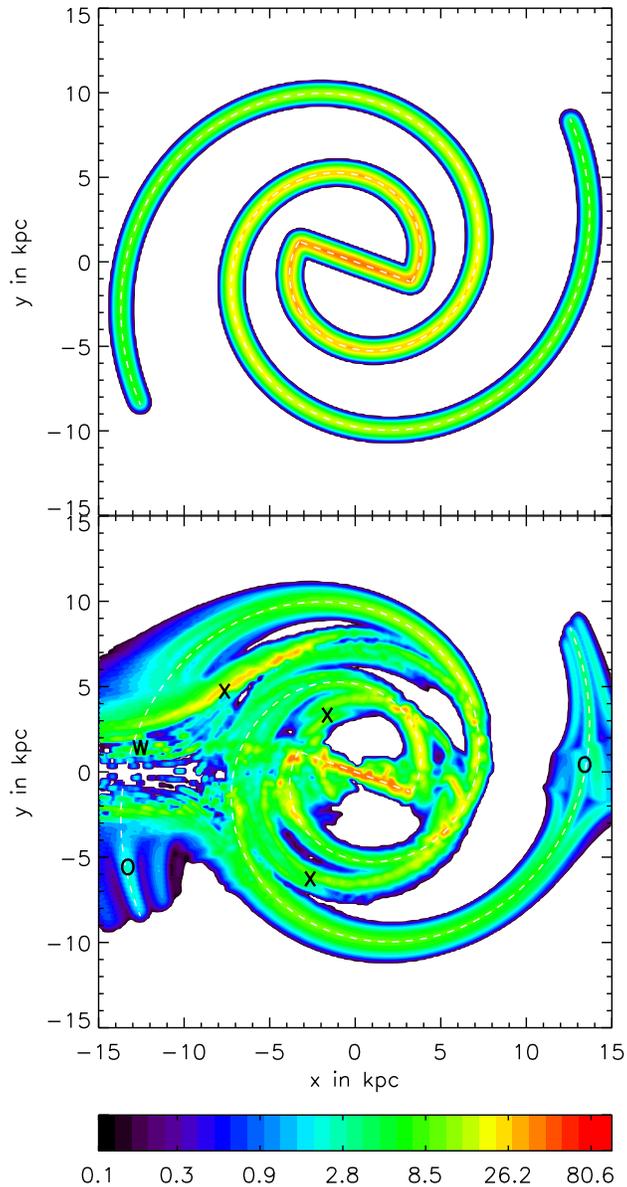}
\caption{(Top panel) The surface mass density in M$_\odot$/pc$^2$ for the simulated gas 
distribution, for which we have calculated hypothetical gas line data.
The dashed white lines illustrate the location of the Galactic bar and 
two possible logarithmic spiral arms, that would correspond to the Sagittarius arm
and the Norma/Perseus +l arm. White areas have a surface mass density below
the lower limit of the colorbar. (Bottom panel) The reconstructed surface mass density
reproduces the bar and the spiral arms with three major types of artefacts, labeled with the letters
{\bf X}, {\bf O}, and {\bf W}.}
\label{f9}
\end{figure}
The bottom panel of figure \ref{f9} shows the reconstructed surface mass density. To be noted
from the figure are three major types of artefacts that we will see again in the deconvolutions of the
real data set. In the regions indicated by the letter {\bf O}, in the anticenter
and outside the solar circle on the far side of the Galaxy, the kinematic resolution is poor or 
nonexistent, and the deconvolution tends to break up the arms into three quasi-parallel structures.
The region labelled  {\bf W} at large negative $x$ has no kinematic resolution, but receives some signal
at zero velocity, which appears here as a tail on the far side of the Galaxy. The signal in the regions
marked with the letter {\bf X} corresponds to the far {distance} solution of gas in the 
spiral arm that passes the sun at about 1 kpc distance. A detailed inspection shows that the 
very high reconstructed surface mass density in the region at $(x,y)=(-8.,5.)$ arises from 
an unusually wide z-distribution. Signal in the wing of the line profile of the highest possible 
velocity near the sun has a velocity, for which only a far solution exists. The deconvolution code 
cannot perfectly separate shift on account of the intrinsic line width from shift arising from the
Galactic gas flow, so a fraction of the signal is attributed to the far solution, which is about 20 times
as distant as the near solution, and so the misplaced signal corresponds to a substantial surface 
mass density, even though it is inconspicuous in the distribution {in distance of the}
gas density.}

\begin{figure}
\epsscale{0.8}
\plotone{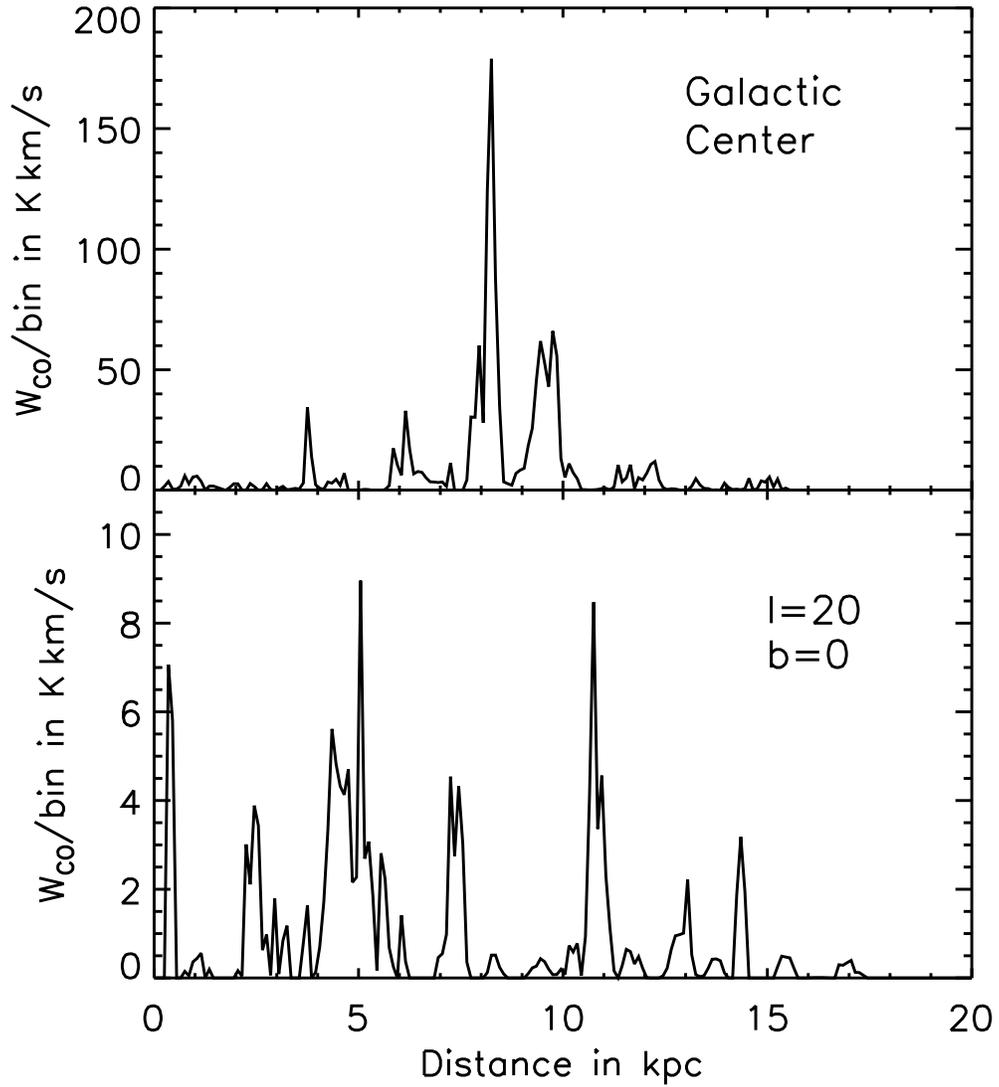}
\caption{The deconvolved integrated CO line intensity per distance bin of 100 pc for the 
Galactic Center direction (top panel) and at 20 degrees positive longitude (bottom). Note the 
difference in scale.}
\label{f3}
\end{figure}
\section{Results}
\subsection{The standard gas-flow model}
Figure \ref{f3} shows the deconvolved gas distribution
(more precisely the integrated line intensity, $W_{\rm CO}$, per distance bin of 100 pc), 
for two lines-of-sight, based on the standard gas-flow model {with 
bar inclination angle $\phi=20^\circ$}.
The top panel refers to the direction of the Galactic Center and can therefore
be directly compared with Figures~\ref{f1} and \ref{f2}, which show the velocity-distance
relation and the CO line spectrum for that line-of-sight. Even though in the
$W_{\rm CO}$ distribution we see a strong narrow peak near the Galactic Center at 8~kpc distance,
a similar fraction of the line signal is in fact placed between about 9 and 10~kpc distance. 
This corresponds to intensity at 50--100~km/s velocity, for which two distance solutions
exist with significantly different $\vert dv/dD\vert$. The two solutions are equivalent 
in terms of the other weight factors, so they receive the same intensity in total, which
in the case of the larger distance is spread out over about ten distance bins.

\begin{figure}
   \plotone{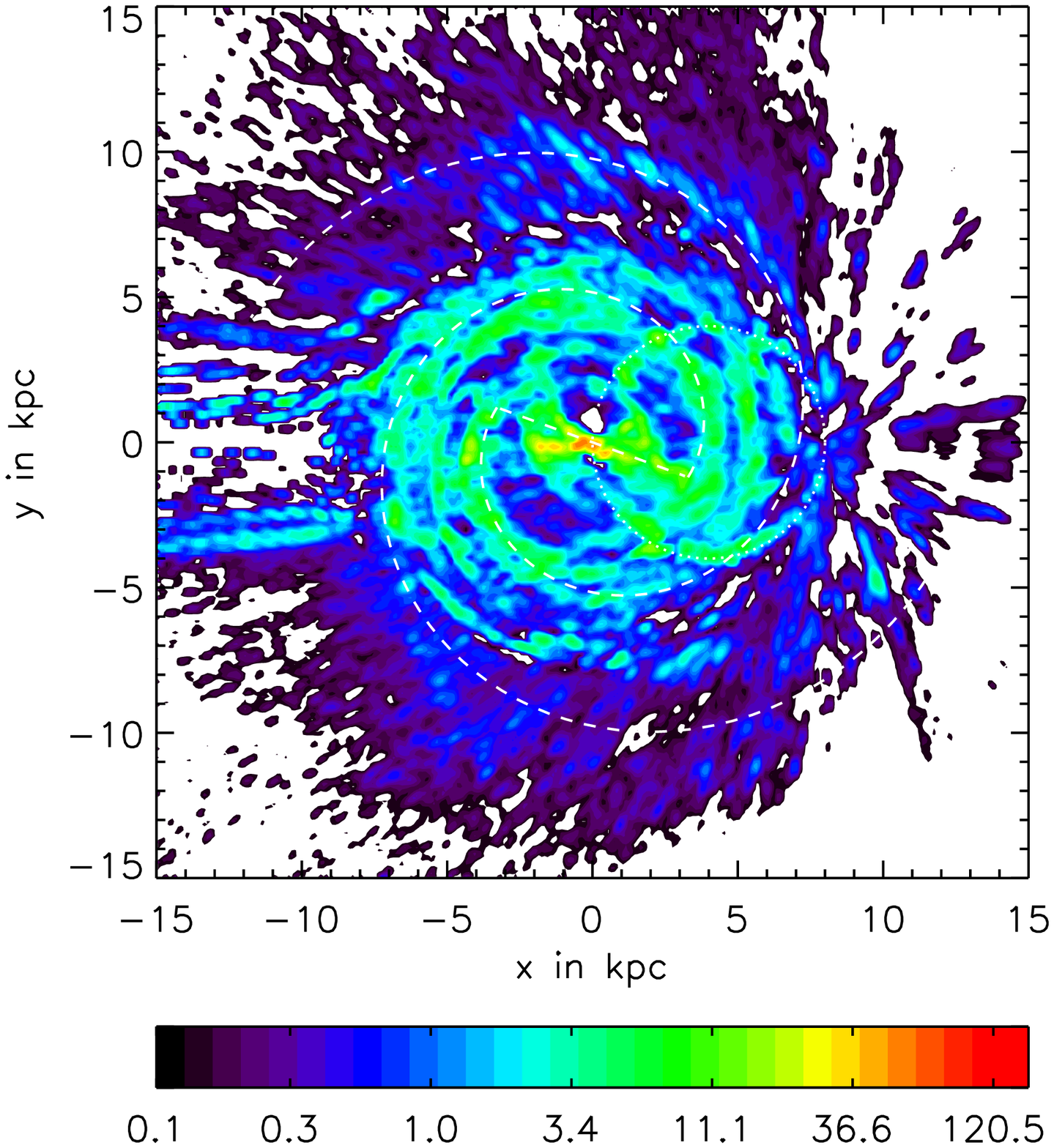}
\caption{The surface mass density of the Galaxy in M$_\odot$/pc$^2$ assuming
a constant conversion factor
$X=2.3\cdot 10^{20}\ {\rm molecules/cm^3/(K\,km\,s^{-1})}$. The dotted white
circle outlines the location of artefacts arising from forbidden velocities.
The dashed white lines illustrate the location of the Galactic bar and 
two possible logarithmic spiral arms, that would correspond to the Sagittarius arm
and the Norma/Perseus +l arm. The data are smoothed to about 200~pc resolution
to better show the large-scale structure. White areas have a surface mass density below
the lower limit of the colorbar.}
\label{f4}
\end{figure}

Figure~\ref{f4} shows the reconstructed surface mass density of molecular gas 
for {the standard gas-flow model and}
a constant conversion factor $X=2.3\cdot 10^{20}\ {\rm molecules/cm^3/(K\,km\,s^{-1})}$
smoothed to about 200~pc resolution. The location of the sun is
$x=8\,{\rm kpc}$ and $y=0$. {The artefact labelled {\bf W} in figure \ref{f9} is 
prominent here as well:} The patches of relatively high surface mass density at large 
Galactocentric radii on the
far side of the Galaxy (at $x\lesssim -10$~kpc) originate from line signal near 
zero velocity, for which those large distances are a valid kinematic solution. In some cases
the line intensity at small velocities is so high, that signal corresponding to a few 
M$_\odot$/pc$^2$ survives the filtering with the Gaussian prior 
that we use to reduce the weights for these solutions.

Also visible in figure~\ref{f4} are artefacts arising from 
gas at forbidden velocities in the inner Galaxy.
Gas at forbidden velocities is placed where the corresponding extremum in the line-of-sight
velocity is found. For circular rotation with constant speed the location of that extremum
{would} delineate a circle of radius $R_0/2$ that extends from the sun to the Galactic Center
and is indicated by the dotted white line. One can clearly 
see patches of high surface mass density that roughly follow the dotted line.
While it should be expected that the gas resides near the location of the peak in the
line-of-sight velocity, the spatial concentration of the gas is most likely exaggerated.
{The fact that a substantial fraction of the total CO line signal is placed near the dotted white
line indicates that the gas flow model underestimates the flow
velocities in the inner Galaxy.}
{Note, that this kind of artefact is not seen in test
  simulation presented in the previous section.}

One clearly sees a mass concentration along the Galactic bar
where the surface mass density is often two orders of magnitude higher than at similar
Galactocentric radii on the sides of the bar. Two spiral arms seem to emerge at the ends 
of the bar. The distribution of molecular gas supports the notion that those two spiral arms 
have a small pitch angle. 
The dashed white line indicates the location of the 
Galactic bar according to \citet{biss03} and, for illustration, two logarithmic
spiral arms with pitch angle $11.5^\circ$, that emanate from the ends of the bar
at $R=3.5$~kpc. On the near side this would be the Norma arm that circles around the
Galactic Center and reappears as the Perseus +l arm in the notation of \citet{vallee}.
Emerging on the far side and closely passing by the sun would be the Sagittarius arm. 
To be noted from the figure is that the distribution of molecular mass is roughly
consistent with those two arms. There is excess molecular material that one may associate with 
two other arms, for example the gas near $x=5$~kpc and $Y=0$~kpc would mark the Scutum arm. 
The gas between the sun and the outer Perseus +l arm would lie in the Perseus arm. While 
certain structures in the map can be associated with those arms as discussed in the literature, 
{it is not a priori clear that those structures are real. A comparison with the 
deconvolved mass distribution for the simulated data set in Fig.~\ref{f9} shows that the excess material
coincides with two artefacts marked by an {\bf X} at $(x,y)=(-2,3)$~kpc and at $(x,y)=(-2,-7)$~kpc,
that correspond to the far solution of the nearby Norma arm and
Sagittarius arm, respectively. On the other hand, the gas distribution in the SPH simulation
of \citet{biss03} does not simply follow the logarithmic spiral arms, in particular not within the
inner few kpc. While the signal at $(x,y)=(-2,-7)$~kpc is close to the boundary of the SPH 
simulation region, {thus hindering a fair assessment of their relevance,}
the structure near $(x,y)=(-2,3)$~kpc
has a clear counterpart in the gas distribution according to the gas model of \citet{biss03}.
}

\begin{figure}
   \plotone{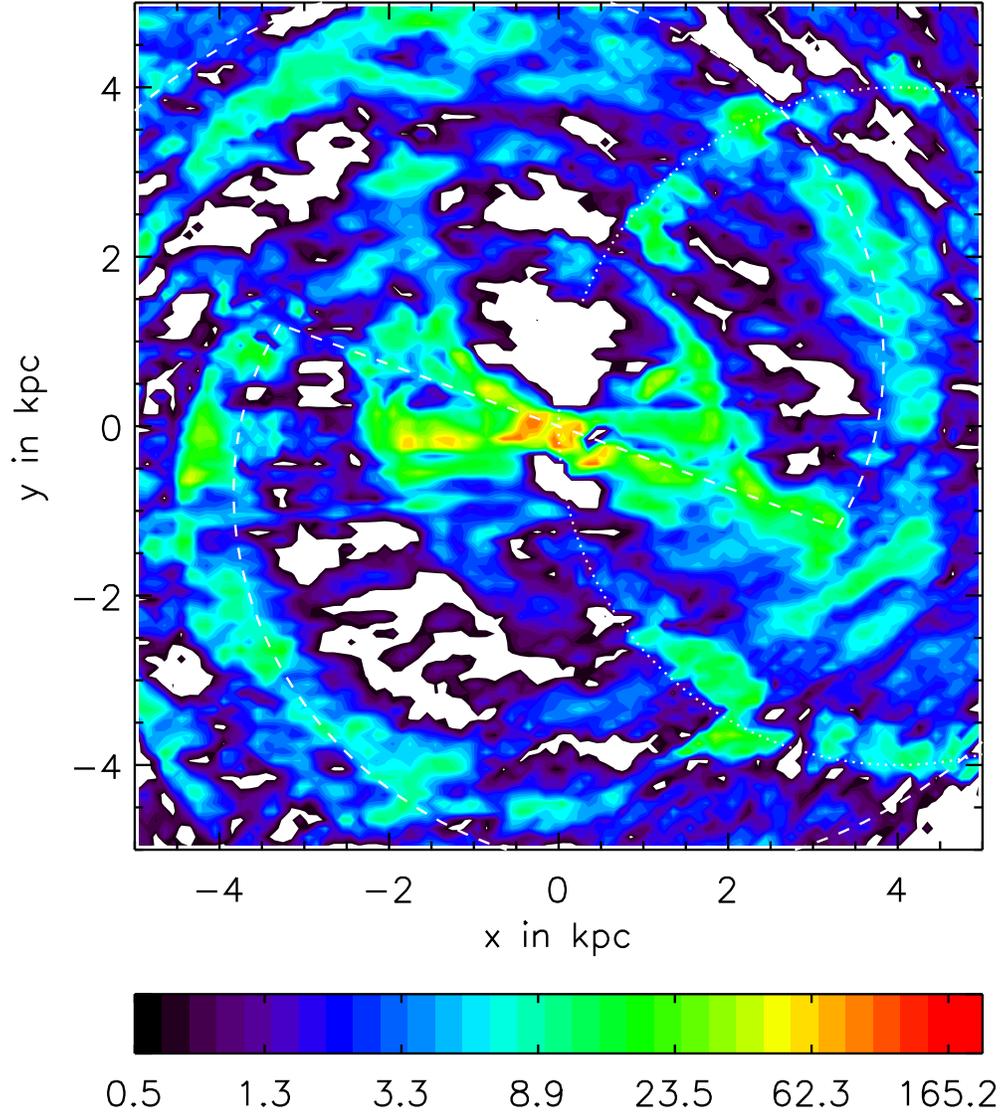}
\caption{The surface mass distribution in the inner 10~kpc $\times$ 10~kpc of the
Galaxy in full resolution. As before the units are
M$_\odot$/pc$^2$ assuming a constant conversion factor
$X=2.3\cdot 10^{20}\ {\rm molecules/cm^3/(K\,km\,s^{-1})}$. White areas have a surface mass density below
the lower limit of the colorbar.}
\label{f5}
\end{figure}

\begin{figure}
   \plotone{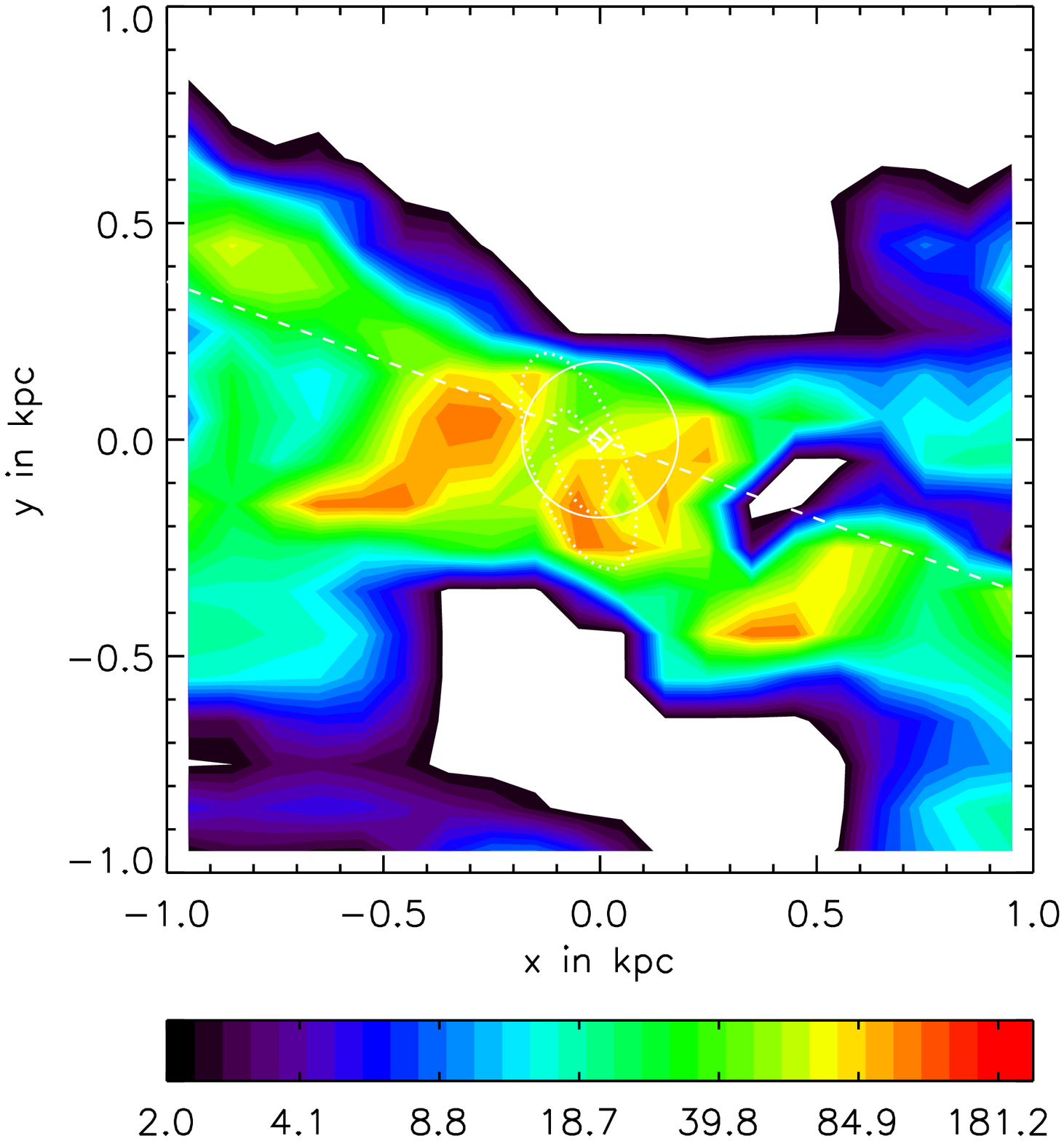}
\caption{Blow-up of the inner 2~kpc $\times$ 2~kpc of Figure~\ref{f5}. The diamond marks the 
position of the Galactic Center. The solid circle outline the massive cold torus structured inferred by
\citet{laun}. The dotted lines indicate the central molecular zone according to the
model of \citet{ferriere}. The inner ellipse
corresponds to the peak in surface mass density, and the outer ellipse outline the perimeter where
the density has fallen to $1/e$ of the peak value. White areas have a surface mass density below
the lower limit of the colorbar.}
\label{f6}
\end{figure}

Figure~\ref{f5} shows the surface-mass distribution within 5~kpc from the Galactic Center
with standard resolution of 100~pc. As before the dashed lines indicate the bar and two 
logarithmic spiral arms, whereas the dotted lines outlines the circle on which gas at 
forbidden velocity may be projected. 
Many of the structures shown in Figs.~\ref{f4} and \ref{f5} do not appear as clean and narrow
as those in the SPH simulations of \citet{biss03}, but the fact that the most salient features 
can be recovered lends credibility to their existence in the Galaxy. {
Particularly interesting are the pseudo-arms that emerge from the bar at about 2~kpc
from the Galactic Center. Our deconvolution shows similar structures at $(x,y)=(2,0)$~kpc
and $(x,y)=(2,0)$~kpc, although a significant fraction of the latter is clearly gas at the
Galactic Center that is misplaced at the far solution (compare Fig.~\ref{f1}).}

Within 1 kpc from the Galactic Center the model velocities are affected by the limited
resolution of the SPH simulation. One may therefore not expect the gas distribution to be well reproduced.
Figure~\ref{f6} shows the reconstructed distribution of molecular gas within 1 kpc of
the Galactic Center. To be noted from the figure is the concentration of molecular gas 
in three segments that may be interpreted as fragments
of an elongated ring that is somewhat off-center, shifted toward positive longitudes 
(negative $y$). The overall geometry resembles
the expanding molecular ring proposed much earlier by \citet{scov72}, but stretched 
along the line-of-sight. {Fig.~\ref{f6} should be interpreted very carefully,}
{because the limited resolution of the gas flow model has a 
significant impact on the reconstructed gas distribution. In fact we find that it depends on how
one interpolates the gas velocity between the grid points at which the average flow 
velocity is computed in the SPH simulation. For comparison we indicate the location of the
massive cold torus found by \citet{laun} using IRAS and COBE/CIRBE data by the
solid white line.
We can further} compare our results with those of
\citet{saw04} who used only observational data (emission vs. absorption) and no kinematic tracing.
They find an center-filled ellipsoidal configuration with large inclination angle to the line-of-sight. 
The peak in surface density appears at $x\approx -0.03$~kpc and $y\approx -0.15$~kpc 
(our coordinate system) and may be identified
with Sgr B and what they denote the 1.3$^\circ$ region. In contrast, we find the peak in mass density
clearly behind the Galactic Center at $x\approx -0.3$~kpc, and the gas distribution is
closer to a ring or two arms as suggested by \citet{sofue}.
\citet{ferriere} have proposed a model of the central molecular zone that
appears to be a compromise between the various configurations discussed in the literature. The
overall appearance of the central molecular zone is that of an ellipse with a slight reduction of 
the density toward the center, which is indicated in figure~\ref{f6} by dotted lines. The inner ellipse
corresponds to the peak in surface mass density, and the outer ellipse outline the perimeter where
the density has fallen to $1/e$ of the peak value. The major axis of the ellipse is nearly perpendicular
to the line-of-sight, whereas in our model it is significantly stretched along the line-of-sight, probably
as a result of the limited resolution of the gas flow model. We find no counterpart to the 
kpc-scale Galactic Bulge disk in the model of \citet{ferriere}, in fact our reconstructed surface 
mass densities to the sides of the bar are very much lower than in her model, if
one accounts for a small X-factor in the Galactic Center region. 
Both in the SPH simulation of \citet{biss03} and in our surface density 
maps one sees a concentration of gas along the bar and large voids to the side, 
which are intersected by pseudo-arms.

\subsection{Alternative gas-flow models}
\begin{figure}
   \plotone{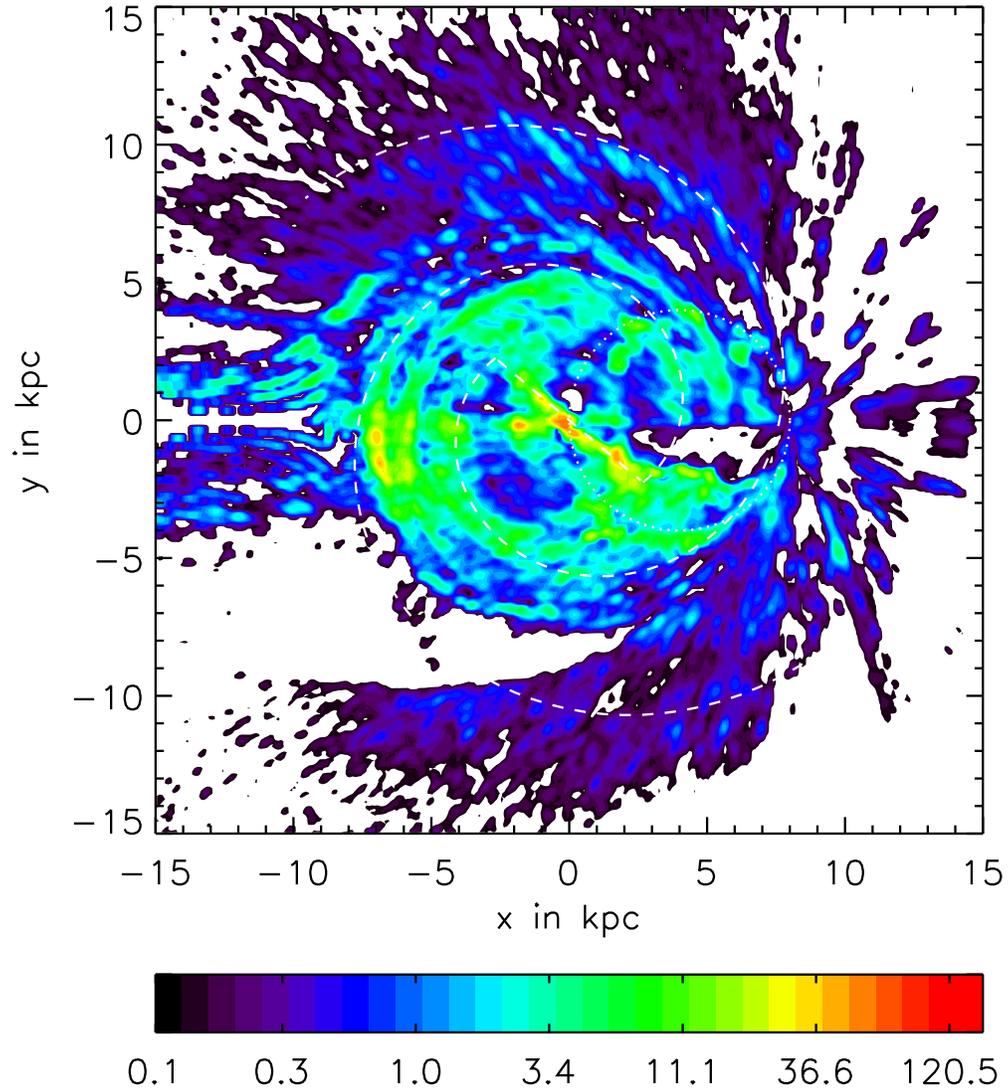}
\caption{The surface mass density in M$_\odot$/pc$^2$ for the standard gas-flow model 
rotated by another $20^\circ$, so that the bar inclination is $\phi=40^\circ$. 
This gas-flow that is intentionally distorted
to no longer reproduce the actual
velocity distribution in the inner Galaxy.}
\label{f11}
\end{figure}

{The gas-flow models are not perfect, for example they may not match both the observed
terminal-velocity curve (TVC) in the inner Galaxy and the orbit velocity at the solar radius to better 
than 5\%.
To investigate the impact of the particular choice of
the gas-flow model on the reconstructed gas distribution, we
here show deconvolutions based on two alternative models. The first is also derived from 
a SPH simulation of \citet{biss03}, but in this case the bar is assumed to be inclined 
at $\phi=30^\circ$. This model should reproduce the terminal velocity curve in the longitude range
$10^\circ \le l\le 60^\circ$, and thus be qualitatively similar to the standard model, except the bar 
parameters are different {and the overall fit of the model to the
  lv-diagram is slightly {worse} but still in agreement with observations.} Figure~\ref{f10} shows 
the reconstructed surface mass density for this
alternative gas-flow model. To be noted from the figure are the large voids at Galactocentric radii
$R=(7-8)$~kpc, whereas the mass density in the inner 6~kpc is more homogeneous than for the 
standard model. Figure~\ref{f10} clearly shows an increased amount of material in
artefacts along the dotted circle, which indicates that the standard
model is more in agreement with the observed gas flow, as expected.

\begin{figure}
   \plotone{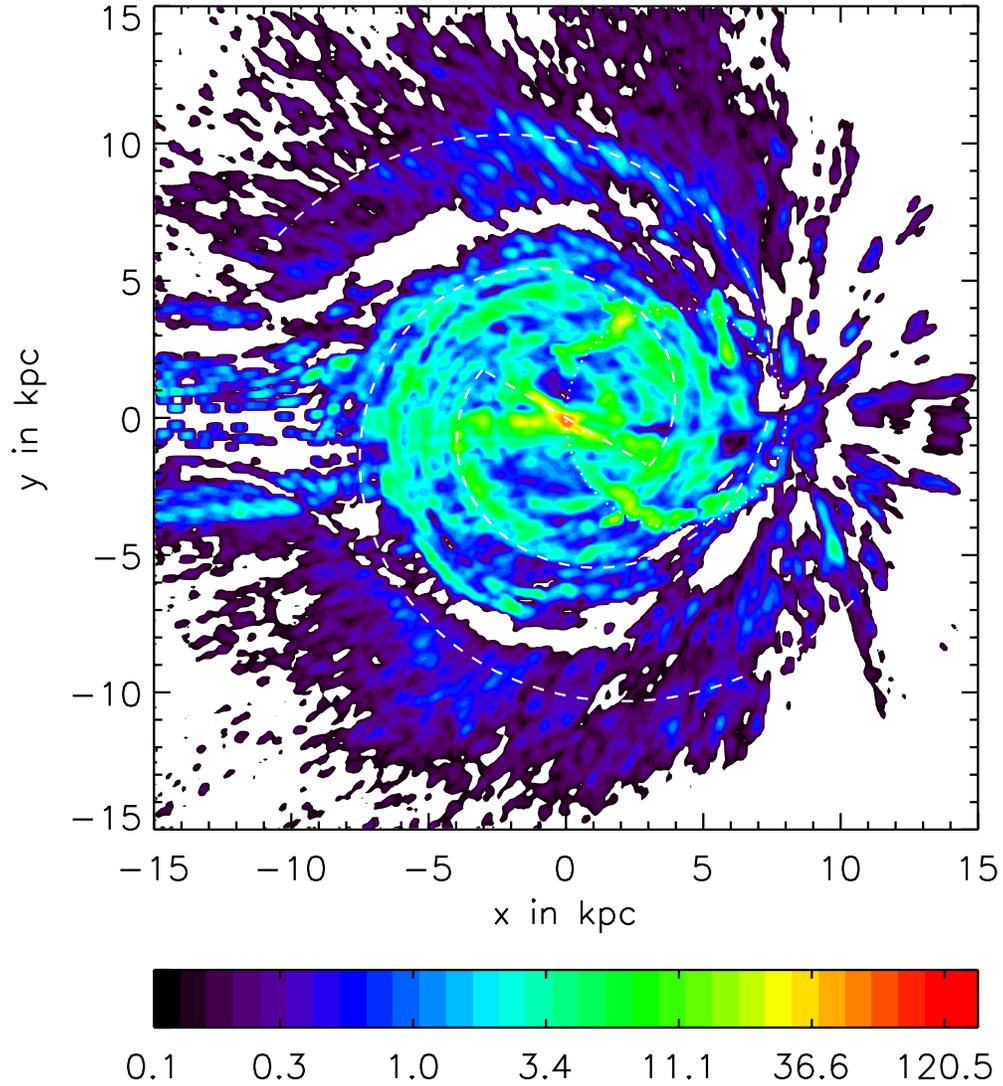}
\caption{The surface mass density in M$_\odot$/pc$^2$ for the alternative gas-flow model 
with bar inclination $\phi=30^\circ$. The dashed line indicating the location of the bar and 
two logarithmic spiral arms is plotted accordingly. White areas have a surface mass density below
the lower limit of the colorbar.}
\label{f10}
\end{figure}

Whereas the first alternative gas-flow model should still be a fair approximation of the actual
velocity distribution in the inner Galaxy, we now} {also} {use a model that is 
intentionally distorted
to no longer reproduce the TVC in the inner Galaxy. For that purpose we rotate the standard gas-flow
model by another $20^\circ$, so that the bar inclination is $\phi=40^\circ$. As shown in Fig.~\ref{f11}, 
the resultant gas distribution is significantly {changed.}}
{There is a large void between the sun and 
the Galactic Center at about $(x,y)=(5,-1)$~kpc, where the line-of-sight velocity in the flow model
no longer matches any signal in the CO line. A second large void appear in the lower left of the plot,
corresponding to a Galactic longitude $l\approx 30^\circ$ and distances around 15~kpc. A relatively
high surface mass density is found on the far side of the Galaxy at 7~kpc Galactocentric 
radius which is most likely misplaced line signal from the Galactic Center region.}

\section{Summary}
We have derived a new model of the distribution of molecular gas in the Galaxy based on 
CO line emission \citep{dame01}. For that purpose we use a gas-flow model derived from 
smoothed particle hydrodynamics (SPH) simulations in gravitational
potentials based on the NIR luminosity distribution of the bulge and disk
\citep{biss03}. Besides providing a more accurate picture of cloud orbits in the inner
Galaxy, {a} fundamental advantage of this model is
that it provides kinematic resolution toward the Galactic Center, in
contrast to standard deconvolution techniques based on purely circular rotation.

{To test the deconvolution procedure and estimate the nature and strength of deconvolution 
artefacts, we have applied it to a simulated CO line data set
based on a model gas distribution that consists of a bar 
and two spiral arms. We have also deconvolved the actual observed CO data using alternative
gas-flow models, one of which is intentionally distorted to no longer reproduce the actual
velocity distribution in the inner Galaxy. The reconstructed distribution of surface mass density 
is significantly affected in the case of the ill-fitting gas-flow model. When using gas-flow models
that reproduce the terminal-velocity curve, but are based on different bar inclination angles, the 
reconstructed gas distribution are much {more alike.}} {In particular, the deconvolution
is robust against a simple rescaling of the gas-flow velocities by a few percent.} 
{A comparison of the surface mass
density determined using feasible and unfeasible gas flow models shows that in the latter
case the resulting surface mass density is also unfeasible. {Examples of that are}
the very low surface mass density
regions around $(x,y) = (5,-1)\,$kpc and $(x,y)=(-8,-8)\,$kpc, and the strong feature at 
$(x,y)=(-7,-1)\,$kpc in Figure~\ref{f11}.
Hence, the comparison of the surface mass densities from the various gas flow models 
strongly indicates that the result of the deconvolution algorithm is robust against 
moderate variations in the underlying gas flow model. On the other hand, it is sensitive enough 
to changes in the gas flow model to discriminate the surface mass density solutions 
based on feasible gas-flow models against those based on unfeasible flow} models. 

{We now describe our model for the surface mass density of the gas in more detail. 
In the model} we find a concentration of mass along the Galactic bar
where the surface mass density is often two orders of magnitude higher than at similar
Galactocentric radii of the sides of the bar. Two spiral arms seem to emerge at the ends 
of the bar, which have a small pitch angle $\sim 12^\circ$. While 
certain structures in the surface density distribution may be associated with two more spiral arms 
as discussed in the literature \citep{vallee}, the evidence for those arms provided by this
deconvolution is not strong, and localizing spiral arms based on kinematics and CO line data
alone is difficult. We also reproduce a concentration of molecular gas 
in the shape of an elongated ring around the Galactic Center that resembles the 
massive cold torus found by \citet{laun}, but is broken up 
and somewhat stretched along the line-of-sight, probably
as a result of the limited resolution of the gas flow model.

Models of the three-dimensional distribution of molecular gas in the 
Milky Way Galaxy can be used in many applications, for example to analyze 
the diffuse Galactic gamma-ray emission that will be observed with
GLAST. Knowledge of the gas distribution is essential for studies of the cosmic-ray
gradient in the Galaxy, but also for investigation of small-scale variations
in the density and flux of cosmic rays. Our gas model will be publicly available at
{\it http://cherenkov.physics.iastate.edu/gas}. 

\acknowledgments
Support by NASA grant NAG5-13559 is gratefully acknowledged.

\end{document}